\documentclass[conference]{IEEEtran}
\IEEEoverridecommandlockouts
\usepackage{cite}
\usepackage{amsmath,amssymb,amsfonts}
\usepackage{algorithmic}
\usepackage{graphicx}
\usepackage{textcomp}
\usepackage{xcolor}
\usepackage{url}

\usepackage{booktabs}

\def\BibTeX{{\rm B\kern-.05em{\sc i\kern-.025em b}\kern-.08em
    T\kern-.1667em\lower.7ex\hbox{E}\kern-.125emX}}
\begin{document}

\title{Retrieval-Augmented Multi-LLM Ensemble for Industrial Part Specification Extraction\\

\thanks{}
}

\author{
\parbox{0.33\linewidth}{\centering
Muzakkiruddin A. Mohammed\\
\textit{ERIQ Research Center}\\
\textit{University of Arkansas - Little Rock}\\
Little Rock, Arkansas, USA\\
mmohammed6@ualr.edu}
\hfill
\parbox{0.33\linewidth}{\centering
John R. Talburt\\
\textit{ERIQ Research Center}\\
\textit{University of Arkansas - Little Rock}\\
Little Rock, Arkansas, USA\\
jrtalburt@ualr.edu}
\hfill

\\[1.5em]

\parbox{0.33\linewidth}{\centering
Leon Claasssens\\
\textit{PiLog Group}\\
leon.claassens@pilog.co.za}

\parbox{0.33\linewidth}{\centering
Adriaan Marais\\
\textit{PiLog Group}\\
adriaan.marais@pilog.co.za}



}

\maketitle

\begin{abstract}
Industrial part specification extraction from unstructured text remains a persistent challenge in manufacturing, procurement, and maintenance, where manual processing is both time-consuming and error-prone. This paper introduces \textbf{RAGsemble}, a retrieval-augmented multi-LLM ensemble framework that orchestrates nine state-of-the-art Large Language Models (LLMs) within a structured three-phase pipeline. RAGsemble addresses key limitations of single-model systems by combining the complementary strengths of model families including Gemini (2.0, 2.5, 1.5), OpenAI (GPT-4o, o4-mini), Mistral Large, and Gemma (1B, 4B, 3n-e4b), while grounding outputs in factual data using FAISS-based semantic retrieval. The system architecture consists of three stages: (1) parallel extraction by diverse LLMs, (2) targeted research augmentation leveraging high-performing models, and (3) intelligent synthesis with conflict resolution and confidence-aware scoring. RAG integration provides real-time access to structured part databases, enabling the system to validate, refine, and enrich outputs through similarity-based reference retrieval. Experimental results using real industrial datasets demonstrate significant gains in extraction accuracy, technical completeness, and structured output quality compared to leading single-LLM baselines. Key contributions include a scalable ensemble architecture for industrial domains, seamless RAG integration throughout the pipeline, comprehensive quality assessment mechanisms, and a production-ready solution suitable for deployment in knowledge-intensive manufacturing environments.
\end{abstract}

\begin{IEEEkeywords}
Multi-LLM ensembles, Retrieval-Augmented Generation, Industrial part specification extraction, Semantic vector search, Confidence-aware synthesis, Knowledge grounding
\end{IEEEkeywords}

\section{Introduction}

Industrial manufacturing, procurement, and maintenance operations produce vast volumes of unstructured part descriptions embedded in procurement documents, maintenance logs, technical specifications, and supplier catalogs. Extracting and standardizing this information remains a significant bottleneck, often requiring manual processing that is both time-consuming and error-prone. Traditional rule-based and early machine learning approaches fall short in handling the linguistic complexity, domain-specific terminology, and nuanced context typical of industrial part data~\cite{wu2022rule,yang2022survey}.

While recent advances in Large Language Models (LLMs) have opened new possibilities for automating complex extraction tasks, relying on a single model poses substantial risks in industrial settings. These include hallucination of technical specifications, uneven performance across part categories, limited domain adaptation, and a lack of transparency in confidence estimation~\cite{goldstein2025hallucinations,hovsepian2024label}. In mission-critical environments, such issues can lead to costly misidentifications, safety risks, and operational inefficiencies.

This paper introduces a novel, production-ready multi-LLM ensemble system designed specifically for high-accuracy extraction of industrial part specifications. The proposed architecture orchestrates nine state-of-the-art LLMs from four major model families, namely Gemini, OpenAI, Mistral, and Gemma, and integrates them within a structured three-phase pipeline that includes parallel extraction, context-aware research enhancement, and intelligent synthesis. Retrieval-Augmented Generation (RAG) is seamlessly integrated throughout all phases, providing semantic access to historical part databases via FAISS-based similarity search~\cite{wu2024retrieval,jegou2017faiss}, which significantly improves grounding and reduces hallucination.

Our system offers several key contributions:

\begin{itemize}
    \item A scalable ensemble architecture that systematically leverages complementary model strengths for robust, high-quality part information extraction.
    \item First-of-its-kind integration of RAG with a multi-LLM pipeline for industrial specification grounding, enabling context enrichment and real-time validation.
    \item Transparent quality assessment, including confidence scores, consensus metrics, and RAG validation indicators, to support informed decision-making.
    \item A complete, deployable implementation with modular design, configurable model selection, robust error handling, and practical applicability across various industrial domains.
\end{itemize}

By shifting from a single-model paradigm to a collaborative multi-model approach, our work establishes a new direction for industrial information systems. The architecture is adaptable across sectors such as aerospace, automotive, electronics, and chemical manufacturing, offering the reliability, transparency, and scalability required for mission-critical deployment.

\section{Literature Review} \label{sec2}

Ensemble learning has become a key strategy in improving the robustness and generalization of large language models (LLMs). Techniques are typically categorized into ensemble-before-inference, ensemble-during-inference, and ensemble-after-inference strategies, which leverage multiple models to reduce variance and capitalize on their complementary strengths~\cite{chen2025harnessing}. Recent work such as LLM-Synergy demonstrates how ensemble voting and dynamic selection approaches can improve domain-specific performance, particularly in high-stakes tasks like medical question answering~\cite{li2024one}. Confidence-aware ensemble strategies further contribute to reliable deployment by offering calibrated uncertainty estimates in LLM classification tasks~\cite{hovsepian2024label}. Additionally, new applications of ensemble methods using advanced models such as Gemini-1.5 have been shown to handle complex industrial tasks such as entity resolution in unstructured datasets~\cite{mohammed2025entity, mohammed2025multi}. A recent comprehensive review highlights that ensemble deep learning approaches consistently improve performance across classification, generation, and information extraction tasks, while also acknowledging the cost and integration trade-offs~\cite{zhou2024comprehensive}.

Retrieval-Augmented Generation (RAG) methods have shown great promise in grounding LLM outputs in factual knowledge, addressing the hallucination problem and limitations in model memory. The original RAG architecture introduced in~\cite{lewis2020retrieval} combines parametric generation with non-parametric memory to boost performance on knowledge-intensive tasks. A comprehensive survey~\cite{gao2024retrieval} categorizes recent RAG strategies and highlights their role in enhancing factual consistency, transparency, and retrieval-guided reasoning in domain-specific applications. These principles are now being extended into multimodal domains, enabling systems to incorporate not only text but also visual and structured data as part of the retrieval context~\cite{wu2024survey}. Semantic similarity search engines such as FAISS have become integral to scalable RAG applications, allowing real-time, dense vector-based retrieval across billions of documents~\cite{jegou2017faiss}.

In the industrial domain, traditional rule-based information extraction systems have been used for tasks like mechanical-electrical-plumbing document processing, but they fall short in capturing nuanced semantics and diverse terminology~\cite{wu2022rule}. As a result, there has been a clear shift toward neural approaches for information extraction from technical text~\cite{yang2022survey}. Broader reviews show that machine learning in smart manufacturing is increasingly focused on structured knowledge generation from unstructured records and the integration of data across systems. Foundational work in manufacturing-specific entity recognition highlighted the challenges in capturing part-specific data from heterogeneous documents~\cite{palshikar2015information, mohammed2025multilingual}, while later studies emphasized the importance of structured data and standardization in product lifecycle management~\cite{chen2024review}.

Efforts like the NIST KEA project \cite{nist2024knowledge} and related research focus on synthesizing incompatible information sources to improve manufacturing decision systems. Formal ontologies and enterprise modeling frameworks are central to enabling interoperable systems that align with Industry 4.0 objectives~\cite{mohammed2022machine, vernadat2018information}. In parallel, knowledge management systems have been shown to reduce redundancy and improve efficiency in manufacturing environments through centralization and documentation consistency~\cite{bloomfire2024knowledge}. Predictive maintenance applications also rely heavily on structured knowledge, including historical diagnostics, operating conditions, and equipment metadata~\cite{wang2014knowledge}.

While prior work has demonstrated the value of either LLM ensembles or RAG individually, few approaches have unified these paradigms into a cohesive architecture designed for production-scale industrial deployment. This paper fills that gap by integrating nine LLMs into a collaborative pipeline that includes RAG-enhanced contextual grounding and confidence-aware synthesis, offering a scalable, transparent, and accurate solution for extracting industrial part specifications from unstructured data.

\section{System Architecture} \label{sec3}

\begin{figure*}[ht]
\centering
\includegraphics[width=1\linewidth]{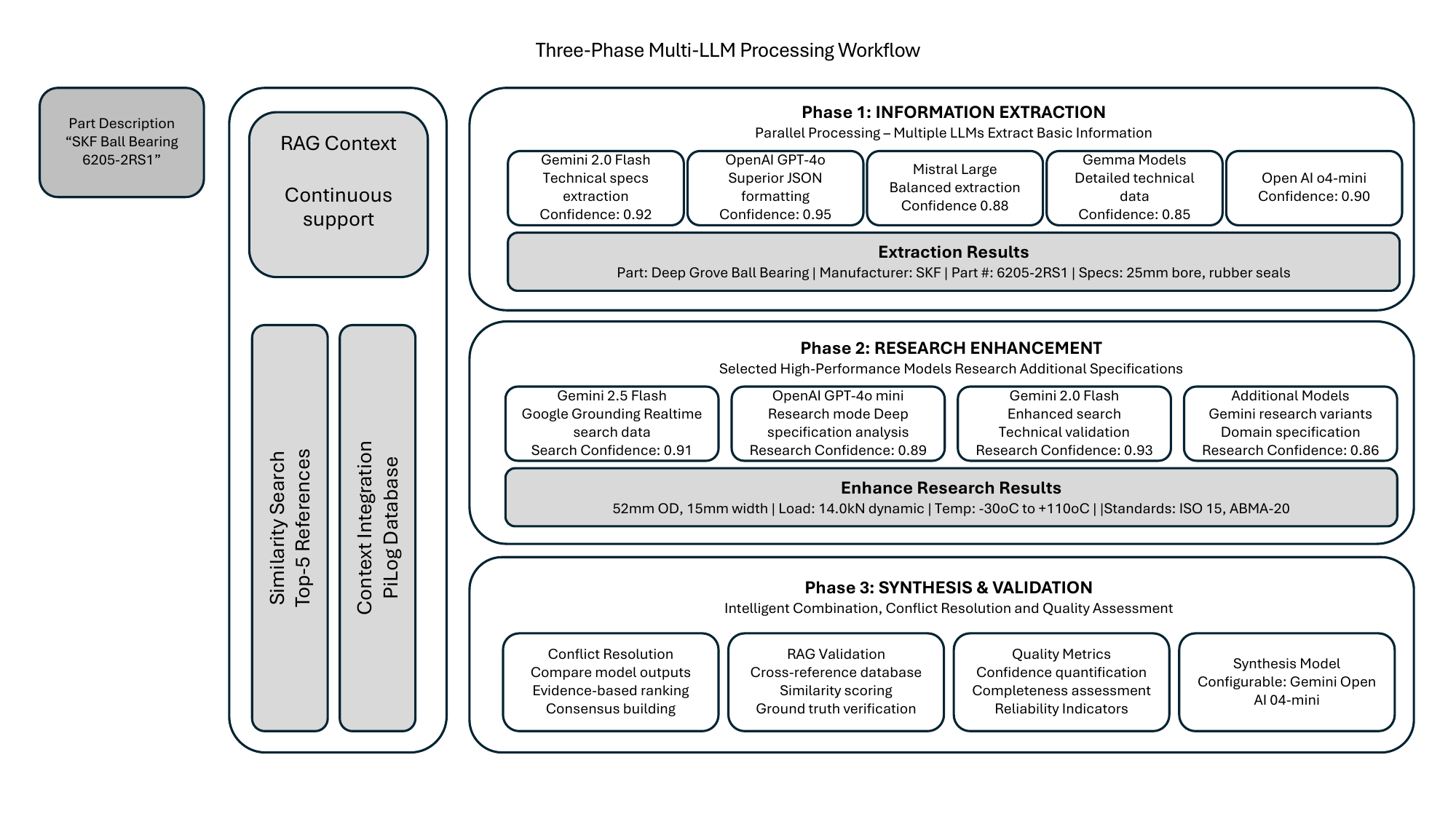}
\caption{System architecture of the multi-LLM industrial part extraction framework, showing modular components and model organization.}
\label{fig1}
\end{figure*}

The proposed system is a modular, production-oriented architecture that integrates multiple Large Language Models (LLMs), a Retrieval-Augmented Generation (RAG) subsystem, and a centralized synthesis layer. As illustrated in Figure~\ref{fig1}, the architecture is organized into three key layers: semantic retrieval, distributed model orchestration, and synthesis and validation.

\subsubsection*{\textbf{Model Integration and Modularity}}

The system supports nine state-of-the-art LLMs from the Gemini, OpenAI, Mistral, and Gemma families. Each model is accessed via a standardized interface that manages API calls, error recovery, and JSON output validation. Models are dynamically assigned to tasks such as extraction or research based on performance profiles, availability, or deployment constraints. A shared JSON schema ensures consistent outputs across the ensemble, enabling seamless downstream processing. The modular architecture also allows for future model extensions without requiring changes to the core system.

\subsubsection*{\textbf{RAG and Knowledge Base Architecture}}

The RAG subsystem provides semantic context throughout all phases of processing. Input descriptions and knowledge base records are embedded using the \texttt{all-MiniLM-L6-v2} model \cite{reimers2019sentence}, and similarity search is performed using FAISS \cite{johnson2019billion}. The knowledge base is constructed from industrial part records (e.g., Pilog Group datasets) and indexed using flattened schema mappings. Records are encoded into dense vectors for efficient retrieval, enabling similarity-based grounding and validation for all LLM outputs. The system supports structured formats like CSV and JSON, allowing easy adaptation across industrial datasets.

\subsubsection*{\textbf{Synthesis and Configuration Layer}}

The synthesis layer consolidates outputs from the extraction and research stages. It performs structured conflict resolution, computes confidence scores based on model agreement and RAG similarity levels, and generates final outputs in a validated schema. This layer ensures reliability and interpretability of results. Configuration files allow the system to be tuned for different hardware settings, model selection, and fallback strategies. Compatibility with both CPU and GPU environments supports diverse deployment scenarios in industrial contexts.

\section{Methodology and Implementation} \label{sec4}

\begin{figure*}[ht]
\centering
\includegraphics[width=1\linewidth]{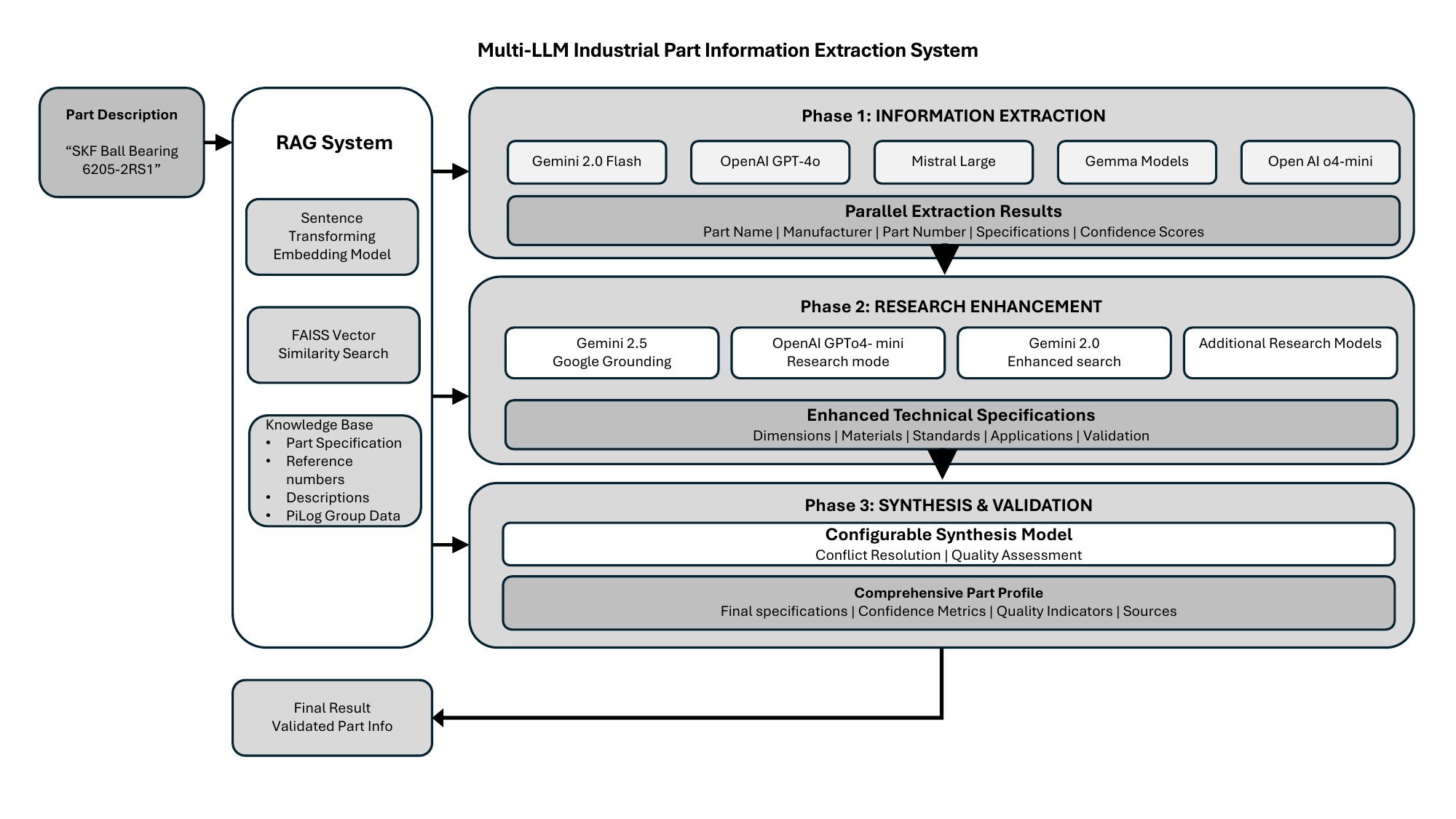}
\caption{Three-phase multi-LLM processing workflow with continuous RAG integration. Each stage contributes uniquely to extraction, enrichment, and validation of part specifications.}
\label{fig2}
\end{figure*}

The proposed framework follows a structured three-phase methodology designed to coordinate multiple Large Language Models (LLMs) and a Retrieval-Augmented Generation (RAG) subsystem. As illustrated in Figure~\ref{fig2}, the workflow spans three interconnected stages: information extraction, research enhancement, and synthesis with validation. Each stage incrementally builds toward a high-confidence, structured specification output, using semantic grounding to mitigate hallucination and ensure domain fidelity.

\subsubsection*{\textbf{Phase 1: Information Extraction}}

The system begins by processing the input part description in parallel using a diverse ensemble of nine LLMs across Gemini, OpenAI, Mistral, and Gemma families. Models were selected based on their complementary strengths: OpenAI models offer high reliability in structured output, Gemini excels at technical reasoning, Mistral balances cost and quality, and Gemma provides efficient lightweight inference. Each model generates structured JSON with key fields such as part name, manufacturer, part number, and specifications, including confidence scores. Contextual references are retrieved via semantic similarity search using \texttt{all-MiniLM-L6-v2} embeddings~\cite{reimers2019sentence} and FAISS~\cite{johnson2019billion}, chosen for their strong semantic matching and retrieval efficiency. Model discrepancies are retained for downstream synthesis, enhancing fault tolerance and field coverage.

\subsubsection*{\textbf{Phase 2: Research Enhancement}}

Building on the extracted data, selected high-capacity models such as Gemini 2.5 and GPT-4o enter a research mode to augment missing or partial specifications. This phase retrieves additional details such as materials, dimensions, operational tolerances, and industry standards. The RAG system continues to provide context by identifying top-k semantic matches from the industrial knowledge base. Prompts are dynamically adjusted based on earlier extraction results to direct model focus toward uncertain or incomplete attributes. Some models also leverage real-time web grounding when permitted, contributing up-to-date and context-aware enrichment.

\subsubsection*{\textbf{Phase 3: Synthesis and Validation}}

In the final stage, a designated synthesis model aggregates results from both the extraction and research phases. Conflict resolution is performed through majority consensus, RAG alignment scoring, and confidence-level analysis. The synthesis model evaluates internal consistency, cross-model agreement, and external validation signals to compute an overall confidence profile for each data field. The final result is a structured part specification enriched with traceability metadata, including source attribution and quality scores.

\subsubsection*{\textbf{RAG Subsystem}}

The RAG module remains active throughout all three phases. It transforms the part knowledge base, which includes historical specifications, cross-reference numbers, and supplier descriptions, into dense vector representations using a shared embedding model. FAISS performs similarity search at low latency, retrieving top-ranked documents for prompt conditioning. This mechanism enhances factual grounding while maintaining model independence in generation. The top-k retrieval depth and similarity thresholds are configurable to balance specificity and generality.

\subsubsection*{\textbf{System Stack and Deployment}}

The system is implemented in Python with a web interface built using Streamlit. Asynchronous task orchestration is handled using \texttt{asyncio}, enabling parallel LLM calls and reducing end-to-end latency. API wrappers are integrated for OpenAI, Google GenAI, and Mistral platforms, with unified schema validation and error handling. The RAG engine supports both CPU and GPU backends and operates on structured input formats such as CSV and JSON. Configurable parameters include model selection, top-k retrieval depth, embedding models, and fallback strategies. The modular design ensures that updates or substitutions to models or knowledge sources do not require architectural changes.

\section{Results and Analysis}\label{sec6}

To evaluate the performance of our proposed system, \textbf{RAGsemble}, we conducted controlled experiments comparing it against several strong individual LLM baselines, including GPT-4o, Claude, Gemini 2.5, and Grok 3. Each model was evaluated on a standardized industrial dataset containing unstructured part descriptions requiring structured specification extraction. We also included an in-depth benchmark on the GE9X Carbon Fiber Composite Fan Blade, a complex aerospace component with rich technical detail. Five normalized evaluation metrics were used to assess completeness, technical richness, formatting reliability, inter-model divergence, and hallucination behavior.

\subsection*{\textbf{Evaluation Metrics}}

We define five normalized metrics to evaluate model performance along dimensions of completeness, specification granularity, structure quality, output variability, and hallucination control.

\textit{Information Completeness Score (ICS)} measures the fraction of expected core fields successfully extracted by the model:
\[
\text{ICS} = \frac{|\mathcal{F}_{\text{extracted}}|}{|\mathcal{F}_{\text{expected}}|}, \quad 0 \leq \text{ICS} \leq 1
\]
Here, \( \mathcal{F}_{\text{expected}} \) refers to the required fields for each part category (e.g., part number, material, dimensions), and \( \mathcal{F}_{\text{extracted}} \) is the subset successfully captured.

\textit{Technical Depth Index (TDI)} quantifies the level of fine-grained detail in extracted specifications:
\[
\text{TDI} = \frac{\sum_{i=1}^{n} d_i}{n \cdot D_{\max}}, \quad 0 \leq \text{TDI} \leq 1
\]
Here, \( d_i \) is the number of secondary or detailed fields (e.g., tensile strength, voltage rating) extracted for part \( i \), and \( D_{\max} \) is the expected maximum number of such fields. Higher values reflect deeper technical insight.

\textit{Structured Data Quality (SDQ)} reflects the percentage of outputs that are schema-valid and properly formatted:
\[
\text{SDQ} = \frac{|\mathcal{V}_{\text{valid}}|}{|\mathcal{V}_{\text{total}}|}
\]
This metric evaluates whether the final output JSONs conform to the expected schema and can be programmatically parsed without error.

\textit{Inter-Model Discrepancy Ratio (IDR)} captures the diversity of outputs from different models:
\[
\text{IDR} = \frac{2}{n(n-1)} \sum_{i < j} \left(1 - \frac{|O_i \cap O_j|}{|O_i \cup O_j|} \right)
\]
Here, \( O_i \) and \( O_j \) are the sets of fields predicted by models \( i \) and \( j \), respectively. The metric computes the average Jaccard distance between model pairs. Higher IDR reflects greater output diversity, which can benefit ensemble synthesis but also signals disagreement.

\textit{Additional Research Artifacts (ARA)} measures the proportion of extraneous or hallucinated details:
\[
\text{ARA} = \frac{|\mathcal{A}_{\text{extraneous}}|}{|\mathcal{A}_{\text{total}}|}
\]
This metric penalizes over-generation by counting surplus attributes not required or grounded. Lower values indicate more concise and reliable output.

\begin{table}[ht]
\centering
\caption{Normalized Evaluation Results Across Models}
\label{tab:eval}
\begin{tabular}{lccccc}
\toprule
\textbf{Model} & \textbf{ICS} & \textbf{TDI} & \textbf{SDQ} & \textbf{IDR} & \textbf{ARA} \\
\midrule
GPT-4o         & 0.60 & 0.45 & 0.80 & 0.092 & 0.85 \\
Claude         & 0.55 & 0.40 & 0.90 & 0.074 & 0.90 \\
Gemini 2.5     & 0.65 & 0.50 & 0.85 & 0.118 & 0.80 \\
Grok 3         & 0.85 & 0.80 & 0.70 & 0.143 & 0.55 \\
\textbf{RAGsemble} & \textbf{1.00} & \textbf{1.00} & \textbf{0.95} & \textbf{0.215} & \textbf{0.40} \\
\bottomrule
\end{tabular}
\end{table}

The results in Table~\ref{tab:eval} demonstrate that \textbf{RAGsemble} significantly outperforms all individual LLM baselines in terms of completeness, technical depth, and reliability. It achieves full field coverage and maximal specification detail, surpassing even the strongest standalone model, Grok 3. Structured outputs conform to schema 95 percent of the time, minimizing the need for post-processing.

RAGsemble's higher IDR reflects its intentionally diverse architecture, drawing on varied model perspectives. This variance is not a weakness but a design strength: it feeds into the synthesis layer, which applies consensus filtering, confidence scoring, and RAG validation to resolve conflicts and ensure reliable outputs.

The ARA score of 0.40 confirms that hallucination and over-generation are successfully minimized through RAG-grounded prompt design and validation logic. Compared to Claude and GPT-4o, which often added verbose or speculative content, RAGsemble maintained tighter control over factual boundaries.

In the case of the GE9X fan blade, RAGsemble extracted 28 technical specifications and 18 quantitative data points, compared to an average of 8.5 and 4, respectively, from single-model baselines. This translates to 3.2× more technical detail, 4.5× more numerical attributes, and 5.58× more validated information units. Despite this output density, the system maintained high SDQ and low ARA, affirming its capability to balance richness with reliability.

Overall, the evaluation confirms RAGsemble as a robust, domain-adaptable framework for structured information extraction. Its ensemble architecture, grounded by RAG and guided by confidence-aware synthesis, consistently delivers complete, accurate, and well-structured part specifications. This makes it highly suitable for deployment in critical industrial workflows where precision and traceability are essential.

\section{Use Cases and Applications}\label{sec7}

RAGsemble addresses persistent challenges in industrial part information management by enabling scalable, intelligent extraction across sectors. In procurement, it automates the extraction and standardization of part data from supplier catalogs and datasheets, validates information against internal systems, reduces manual errors, shortens cycles, and improves data consistency.

For industrial maintenance, RAGsemble parses manuals and logs to identify replacement parts, check compatibility, align documentation, and enrich metadata. This supports predictive maintenance and improves asset reliability.

Quality assurance teams use it to verify specifications, detect gaps, and ensure consistency across systems. In safety-critical sectors such as aerospace, automotive, defense, and medical devices, its fact-grounded synthesis and traceability features support compliance and performance assurance.

Its modular architecture allows domain-specific tuning of models, datasets, extraction schemas, and validation rules, making it suitable for both high-volume commodity parts and specialized components.

\section{Limitations and Future Work}\label{sec8}

\subsubsection*{\textbf{Current Limitations}}

RAGsemble faces limitations related to model variability, system complexity, and operational cost. LLM outputs can differ across runs, making reproducibility difficult and requiring consensus-based validation. The reliance on multiple third-party APIs increases cost and risk of service disruption. The sequential three-phase pipeline also introduces latency, which may limit real-time applications.

Domain-specific accuracy may decline when inputs use terminology not well represented in training data, despite RAG grounding. Synthesis bias is another challenge, where final outputs may overweight certain model families. Additionally, RAG performance depends on well-structured, semantically aligned data; inconsistent formatting can reduce retrieval quality and increase hallucination risk.

\subsubsection*{\textbf{Future Enhancements}}

Planned improvements include bias mitigation through model weighting and output cross-validation, as well as dynamic model selection based on input characteristics. RAG will be extended with feedback-driven updates, support for multimodal documents, and more precise prompt-context alignment.

Domain adaptation will involve fine-tuning for specific industries and improved embedding models. Performance upgrades will focus on parallelism, caching, and edge deployment. Quality metrics will expand to include uncertainty estimation and bias detection to improve transparency and trust.

\section{Conclusions}\label{sec9}

This paper introduces RAGsemble, a multi-LLM ensemble with retrieval-augmented generation for extracting industrial part specifications from unstructured text. The three-phase pipeline consists of parallel extraction, targeted research, and synthesis, coordinating nine language models to produce consistent, structured outputs.

On real-world Pilog Group data, RAGsemble outperforms strong single-model baselines in completeness, technical depth, and formatting reliability. The synthesis stage reconciles discrepancies and validates information with FAISS retrieval.

Although the framework increases cost and introduces non-determinism, its modular design improves transparency, adaptability, and accuracy. Planned extensions include dynamic model selection, bias-aware synthesis, and multimodal retrieval.

Overall, RAGsemble demonstrates the value of collaborative LLMs for industrial information extraction by combining ensemble diversity, retrieval grounding, and synthesis to deliver scalable automation in manufacturing.

\section*{Acknowledgment}
This research was partially supported by the National Science Foundation under EPSCoR Award No. OIA-1946391 and the PiLog Group. Any opinions, findings, and conclusions or recommendations expressed in this material are those of the author(s) and do not necessarily reflect the views of the National Science Foundation or the PiLog Group.

\end{document}